\documentclass[12pt]{article}
\usepackage{amsmath,amssymb,graphicx,calc,capt-of,ifthen,theorem}

\begin{document}
\title{Cosmology in Delta-Gravity.}
\author{Jorge Alfaro and Pablo Gonz\'alez\\
Pontificia Universidad Cat\'olica de Chile, Av. Vicu\~na Mackenna 4860, Santiago, Chile\\}

\maketitle

\begin{abstract}
We present a model of the gravitational field based on two symmetric tensors. Gravity is affected by the new field, but outside matter the predictions of the model coincide exactly with general relativity, so all classical tests are satisfied. We find that massive particles do not follow a geodesic while massless particles trajectories are null geodesics of an effective metric. We study the Cosmological case, where we get an accelerated expansion of the universe without dark energy. We also introduce the possibility to explain dark matter with $\tilde{\delta}$ gravity.
\end{abstract}

\maketitle

\section*{INTRODUCTION.}

We know that general relativity (GR) works very well at the macroscopic scales \cite{GR scale}. Nevertheles, its quantization has proved to be difficult, though. The theory is non-renormalizable, which prevents its unification with the other forces of nature. Trying to make sense of quantum GR is the main physical motivation of string theories \cite{string 1,string 2}. Moreover, recent discoveries in cosmology \cite{Weinberg}-\cite{DM DE 4} have revealed that most part of matter is in the form of unknown matter, dark matter (DM), and that the dynamics of the expansion of the Universe is governed by a mysterious component that accelerates the expansion, dark energy (DE). Although GR is able to accommodate both DM and DE, the interpretation of the dark sector in terms of fundamental theories of elementary particles is problematic \cite{DM DE 5}. Although some candidates exists that could play the role of DM, none have been detected yet. Also, an alternative explanation based on the modification of the dynamics for small accelerations cannot be ruled out \cite{DM DE 6, DM DE 7}.\\

In GR, DE can be explained if a small cosmological constant ($\Lambda$) is present. In early times, this constant is irrelevant, but at the later stages of the evolution of the Universe $\Lambda$ will dominate the expansion, explaining the acceleration. Such small $\Lambda$ is very difficult to generate in quantum field theory (QFT) models, because $\Lambda$ is the vacuum energy, which is usually very large.\\

One of the most important mysteries in cosmology and cosmic structure formation is to understand the nature of dark energy in the context of a fundamental physical theory \cite{DE 1, DE 2}. In recent years there has been various proposals to explain the observed acceleration of the universe. They involve the inclusion of some additional fields in approaches like quintessence, chameleon, vector dark energy or massive gravity; The addition of higher order terms in the Einsten-Hilbert action, like $f(R)$ theories and Gauss-Bonnet terms and finally the introduction of extra dimensions for a modification of gravity on large scales (See \cite{DE 3}).\\

Less widely explored, but interesting possibilities, are the search for non-trivial ultraviolet fixed points in gravity (asymptotic safety \cite{GR Weinberg}) and the notion of induced gravity \cite{induced gravity 1}-\cite{induced gravity 4}. The first possibility uses exact renormalization-group techniques \cite{Ren group 1}-\cite{Ren group 4} together with lattice and numerical techniques such as Lorentzian triangulation analysis \cite{Lorentz triang}. Induced gravity proposes that gravitation is a residual force produced by other interactions.\\

Recently, in \cite{Alfaro 0, Alfaro 1} a field theory model explores the emergence of geometry by the spontaneous symmetry breaking of a larger symmetry where the metric is absent. Previous work in this direction can be found in \cite{work 1}-\cite{work 7}.\\

In this paper, we present a model of gravitation that is very similar to classical GR, but could make sense at the quantum level. In the construction, we consider two different points. The first is that GR is finite on shell at one loop \cite{tHooft}, so renormalization is not necessary at this level. The second is a type of gauge theories, $\tilde{\delta}$ gauge theories (DGT),  presented in \cite{Alfaro 2,Alfaro 3}, which main properties are: (a) New kind of fields are created, $\tilde{\phi}_I$, from the originals $\phi_I$. (b) The classical equations of motion of $\phi_I$ are satisfied in the full quantum theory. (c) The model lives at one loop. (d) The action is obtained through the extension of the original gauge symmetry of the model, introducing an extra symmetry that we call $\tilde{\delta}$ symmetry, since it is formally obtained as the variation of the original symmetry. When we apply this prescription to GR we obtain $\tilde{\delta}$ gravity. Quantization of $\tilde{\delta}$ gravity is discussed in \cite{delta gravity}.\\

Here, we study the classical effects of $\tilde{\delta}$ gravity at the cosmological level. For this, we assume that the universe only has two kind of components, non relativistic matter (DM, baryonic matter) and radiation (photons, massless particles), which satisfy a fluid-like equation $p = \omega \rho$. We will not include the matter dynamics, except by demanding that the energy-momentum tensor of the matter fluid is covariantly conserved. This is required in order to respect the symmetries of the model. In contrast to \cite{DG DE}, where an approximation is discussed, in this work we find the exact solution of the equations corresponding to the above suppositions. This solution is used to fit the supernova data and we obtain a physical reason for the accelerated expansion of the universe within the model: the existence of massless particles. If massless particles were absent, the expansion of the Universe would be the same as in GR without a cosmological constant. In the Conclusions we speculate on a possible physical mechanism that could stop the accelerated expansion and prevent the appearance of a Big Rip.\\

We also introduce a possible explanation of DM with $\tilde{\delta}$ gravity. This proposal looks promising, but a more detailed work is required to fit the velocity curves of stars around the center of galaxies, for example.\\

Therefore, in the cosmological regime, we can say that the main properties of this model at the classical level are: (a) It agrees with GR, far from the sources and with adequate boundary conditions. In particular, the causal structure of $\tilde{\delta}$ gravity in vacuum is the same as in general relativity. (b) When we study the evolution of the Universe, it predicts acceleration without a cosmological constant or additional scalar fields. The Universe ends in a Big Rip, similar to the scenario considered in \cite{Phantom 1}. (c) The scale factor agrees with the standard cosmology at early times and show acceleration only at later times. Therefore we expect that density perturbations should not have large corrections. (d) Apparently, the dark matter could be explain with $\tilde{\delta}$ gravity.\\

It was noticed in \cite{Alfaro 3} that the Hamiltonian of delta models is not bounded from below. Phantoms cosmological models \cite{Phantom 1,Phantom 2} also have this property. Although it is not clear whether this problem will subsist or not in a diffeomorphism-invariant model as $\tilde{\delta}$ gravity. Phantom fields are used to explain the expansion of the universe. So, even if it could be said that our model works on similar grounds, the accelerated expansion of the universe is really produced by a reduced quantity of a radiation component in the universe, not by a phantom field.\\

It should be remarked that $\tilde{\delta}$ gravity is not a metric model of gravity because massive particles do not move on geodesics. Only massless particles move on null geodesics of a linear combination of both tensor fields.\\

\section{\label{Chap: delta Gravity Action and Equation of Motion.}$\tilde{\delta}$ Gravity Action and Equation of Motion.}

Our modified model is based on a given action $S_0[\phi]$, where $\phi_I$ are generic fields, to which we add a new term equal to a $\tilde{\delta}$ variation with respect to the the original fields. We define $\tilde{\delta} \phi_I = \tilde{\phi}_I$ so that we have:

\begin{eqnarray}
\label{Mod Action}
S[\phi,\tilde{\phi}] = S_0[\phi] + \kappa_2 \int \frac{\delta S_0}{\delta \phi_I(x)}[\phi] \tilde{\phi}_I(x)
\end{eqnarray}

where $\kappa_2$ is an arbitrary, but small, constant and the index $I$ refers to any kind of indices. This new action shows the standard structure which is used to define any modified element or function for $\tilde{\delta}$ type models. In particular, let us consider the Einstein-Hilbert Action. This action involves:

\begin{eqnarray}
\label{GR Action}
S_0[g] &=& \int d^4x \sqrt{-g}\left(-\frac{1}{2 \kappa}R + L_M\right) \\
\label{delta GR Action}
S[g] &=& \int d^4x \sqrt{-g}\left(-\frac{1}{2 \kappa}R + L_M + \kappa'_2\left(G^{\mu \nu} - \kappa T^{\mu \nu}\right)\tilde{g}_{\mu \nu} + \frac{\kappa_2}{2}\left(\lambda_{\mu ; \nu}+\lambda_{\nu ; \mu}\right)T^{\mu \nu}\right)
\end{eqnarray}

with $\kappa = \frac{8\pi G}{c^2}$, $\kappa'_2 = \frac{\kappa_2}{2\kappa}$, $L_M$ some matter Lagrangian and where:

\begin{eqnarray}
T^{\mu \nu} = - 2\frac{\delta L_M}{\delta g_{\mu \nu}} - g^{\mu \nu}L_M
\end{eqnarray}

is the energy momentum tensor. The last term in (\ref{delta GR Action}), depending upon the auxiliary fields $\lambda_{\mu}$, is needed to impose the condition $T^{\mu \nu}_{~~;\nu} = 0$ as an equation of motion in order to implement the $\tilde{\delta}$ symmetry (\ref{delta symmetry}) off shell. This term is null in vacuum. So, we have a gravitation model with two fields, the first is just the usual gravitational field $g_{\mu \nu}$ and a second one $\tilde{g}_{\mu \nu} = \tilde{\delta} g_{\mu \nu}$. These fields have the following transformations:

\begin{eqnarray}
\label{delta symmetry}
\delta g_{\mu \nu} &=& \xi_{0 \mu ; \nu} + \xi_{0 \nu ; \mu} \nonumber \\
\delta \tilde{g}_{\mu \nu} &=& \xi_{1 \mu ; \nu} + \xi_{1
\nu ; \mu} + \tilde{g}_{\mu \rho} \xi_{0, \nu}^{\rho} +
\tilde{g}_{\nu \rho} \xi^{\rho}_{0, \mu} + \tilde{g}_{\mu \nu,
\rho} \xi_0^{\rho} \nonumber \\
\delta \lambda_{\mu} &=& \xi_{1 \mu} + \lambda_{\rho}\xi_{0, \mu}^{\rho} + \lambda_{\mu, \rho}\xi_{0}^{\rho}
\end{eqnarray}

where $\xi^{\mu}_0$ and $\xi^{\mu}_1$ are infinitesimal contravariant vectors. The first transformation is the typical general coordinate transformations and the second one is obtained using $\tilde{\delta}$ on the first, with $\xi_1^{\mu} \equiv \tilde{\delta} \xi_0^{\mu}$. The modified action given by (\ref{delta GR Action}) was constructed such that is invariant under these transformations, so they are the starting point of our model.\\

Now, if we vary the fields $\lambda_{\mu}$, $\tilde{g}_{\mu \nu}$ and $g_{\mu \nu}$ in (\ref{delta GR Action}), we obtain respectively the equations of motion:

\begin{eqnarray}
\label{Conserv Eq} T^{\mu \nu}_{~~;\nu} &=& 0 \\
\label{Einst Eq} G^{\mu \nu} &=& \kappa T^{\mu \nu} \\
\label{tilde Eq} F^{(\mu \nu) (\alpha \beta) \rho
\lambda} D_{\rho} D_{\lambda} \tilde{g}_{\alpha \beta} &=& \kappa
\frac{\delta T_{\alpha \beta}}{\delta g_{\mu \nu}}
\left(\tilde{g}^{\alpha \beta} - \lambda^{\alpha ; \beta} - \lambda^{\beta ; \alpha}\right) - \frac{1}{2}\left(R^{\mu \nu}\tilde{g}^{\sigma}_{\sigma} - R
\tilde{g}^{\mu \nu}\right) \\ 
&+& \kappa \left(T^{\mu \nu}_{~~;\alpha}\lambda^{\alpha} + T^{\mu \nu}\lambda^{\alpha}_{; \alpha} + T^{\mu \alpha}\lambda_{\alpha}^{~; \nu} + T^{\nu \alpha}\lambda_{\alpha}^{~; \mu} - g^{\mu \nu} T^{\alpha \beta}\lambda_{\alpha ; \beta}\right)  \nonumber
\end{eqnarray}

with:

\begin{eqnarray}
\label{F} F^{(\mu \nu) (\alpha \beta) \rho \lambda} &=& P^{((\rho
\mu) (\alpha \beta))}g^{\nu \lambda} + P^{((\rho \nu) (\alpha
\beta))}g^{\mu \lambda} - P^{((\mu \nu) (\alpha \beta))}g^{\rho
\lambda} - P^{((\rho \lambda) (\alpha \beta))}g^{\mu \nu} \nonumber \\
P^{((\alpha \beta)(\mu \nu))} &=& \frac{1}{4}\left(g^{\alpha
\mu}g^{\beta \nu} + g^{\alpha \nu}g^{\beta \mu} - g^{\alpha
\beta}g^{\mu \nu}\right)
\end{eqnarray}

where $(\mu \nu)$ denotes that the $\mu$ and $\nu$ are in a totally symmetric combination. We can see that the Einstein's equation do not change. In the application to cosmology, we assume the presence of a perfect fluid, so $T_{\alpha \beta} = p g_{\alpha \beta} + (p + \rho) U_{\alpha}U_{\beta}$ with $p$ and $\rho$ the pressure and density respectively and they are given by (\ref{Conserv Eq}). $U^{\alpha}$ is the four-velocity of the fluid and fulfills the identity $U_{\alpha}U^{\alpha}=-1$. Using this energy-momentum tensor we get:

\begin{eqnarray}
\label{T tilde}
\frac{\delta T_{\alpha \beta}}{\delta g_{\mu \nu}}S^{\alpha \beta} = \frac{1}{2} g_{\alpha \beta} \left(T^{\mu \alpha}S^{\nu \beta} + T^{\nu \alpha}S^{\mu \beta}\right)
\end{eqnarray}

with $S^{\alpha \beta} = \tilde{g}^{\alpha \beta} - \lambda^{\alpha ; \beta} - \lambda^{\beta ; \alpha}$. Using (\ref{Einst Eq}) and (\ref{T tilde}), (\ref{tilde Eq}) is reduced to:

\begin{eqnarray}
\label{tilde Eq 2} F^{(\mu \nu) (\alpha \beta) \rho
\lambda} D_{\rho} D_{\lambda} \tilde{g}_{\alpha \beta} &=& G^{\mu \nu}_{~~;\alpha}\lambda^{\alpha} + R^{\mu \nu}\lambda^{\alpha}_{; \alpha} - g^{\mu \nu} R^{\alpha \beta}\lambda_{\alpha ; \beta} - \frac{1}{2}\left(R^{\mu \nu}\tilde{g}^{\sigma}_{\sigma} - R^{\mu \sigma}\tilde{g}^{\nu}_{\sigma} - R^{\sigma \nu}\tilde{g}^{\mu}_{\sigma}\right) \\ &+& \frac{1}{2}\left(R^{\mu \alpha}\lambda_{\alpha}^{~; \nu} + R^{\nu \alpha}\lambda_{\alpha}^{~; \mu} - R^{\mu \alpha}\lambda^{\nu}_{~; \alpha} - R^{\nu \alpha}\lambda^{\mu}_{~; \alpha}\right) \nonumber
\end{eqnarray}

Since $\lambda_{\mu}$ is a auxiliary field to implement that $T^{\mu \nu}$ is conserved, we can fix it. So, from now we will use the gauge $\lambda_{\mu} = 0$. This gauge preserves general coordinate transformations, but fixes completely the extra symmetry such that $\xi_{1 \mu} = 0$. Finally, the $\tilde{g}_{\mu \nu}$ equation is:

\begin{eqnarray}
\label{tilde Eq 3} F^{(\mu \nu) (\alpha \beta) \rho
\lambda} D_{\rho} D_{\lambda} \tilde{g}_{\alpha \beta} + \frac{1}{2}\left(R^{\mu \nu}\tilde{g}^{\sigma}_{\sigma} - R^{\mu \sigma}\tilde{g}^{\nu}_{\sigma} - R^{\sigma \nu}\tilde{g}^{\mu}_{\sigma}\right) = 0
\end{eqnarray}

With this equation we can completely determine $\tilde{g}_{\mu \nu}$ if we know $g_{\mu \nu}$. Because, outside the sources ($T_{\mu \nu} = 0$), $\tilde{g}_{\alpha \beta} \propto g_{\alpha \beta}$ is a solution to (\ref{tilde Eq 2}) and we have $g_{\alpha \beta} = \eta_{\alpha \beta}$  in the vacuum, our boundary conditions will be $g_{\alpha \beta} = \eta_{\alpha \beta}$ and $\tilde{g}_{\alpha \beta} = \eta_{\alpha \beta}$  far from the sources.\\

In order to explore the phenomenology of the model, we need to know the equation of motion of a test particle. We discuss this in the next section.\\

\section{\label{Chap: delta Test Particle Action.}$\tilde{\delta}$ Test Particle Action.}

To find the coupling of a test particle to a gravitational background field, we follow the prescription to construct $\tilde{\delta}$ models (\ref{Mod Action}). We know that, in the standard case, the test particle action is:

\begin{eqnarray}
\label{Geo Action 0} S_0[\dot{x},g] = - m \int dt \sqrt{-g_{\mu
\nu}\dot{x}^{\mu}\dot{x}^{\nu}}
\end{eqnarray}

Since in our model the modified action is obtained according to (\ref{Mod Action}), where in this case $\phi_{I} = g_{\mu \nu}$, the new test particle action is:

\begin{eqnarray}
\label{Geo Action} S[\dot{x},g,\tilde{g}] = m \int dt
\frac{\bar{g}_{\mu
\nu}\dot{x}^{\mu}\dot{x}^{\nu}}{\sqrt{-g_{\alpha
\beta}\dot{x}^{\alpha}\dot{x}^{\beta}}}
\end{eqnarray}

where $\bar{g}_{\mu \nu} = g_{\mu \nu} + \frac{\kappa_2}{2} \tilde{g}_{\mu \nu}$. But (\ref{Geo Action 0}) and (\ref{Geo Action}) are useless for massless particles. To solve this problem, we will start from:

\begin{eqnarray}
\label{Geo Action 0 Lagr} S_0[\dot{x},g,v] = \frac{1}{2} \int dt
\left(vm^2 - v^{-1}g_{\mu \nu}\dot{x}^{\mu}\dot{x}^{\nu}\right)
\end{eqnarray}

where $v$ is a Lagrange multiplier. This action is invariant under reparametrizations:

\begin{eqnarray}
\label{reparametr} \bar{x}^{\mu}(\bar{t}) &=& x^{\mu}(t) \nonumber \\
\bar{v}(\bar{t}) d\bar{t} &=& v(t) dt \nonumber \\
\bar{t} &=& t -\epsilon(t)
\end{eqnarray}

From (\ref{Geo Action 0 Lagr}), we can obtain the equation of motion for $v$:

\begin{eqnarray}
\label{v eq} v = - \frac{\sqrt{-g_{\mu
\nu}\dot{x}^{\mu}\dot{x}^{\nu}}}{m}
\end{eqnarray}

If we substitute (\ref{v eq}) in (\ref{Geo Action 0 Lagr}), we recover (\ref{Geo Action 0}). In other words, (\ref{Geo Action 0 Lagr}) is a good action that include the massless case. If we substitute (\ref{Geo Action 0 Lagr}) in (\ref{Mod Action}) now, we obtain:

\begin{eqnarray}
\label{Geo Action Lagr 1} S[\dot{x},g,\tilde{g},v,\tilde{v}] =
\frac{1}{2} \int dt \left(vm^2 - v^{-1}\left(g_{\mu \nu} +
\kappa_2 \tilde{g}_{\mu \nu}\right)\dot{x}^{\mu}\dot{x}^{\nu} +
\kappa_2 \tilde{v} \left(m^2 + v^{-2} g_{\mu
\nu}\dot{x}^{\mu}\dot{x}^{\nu}\right)\right)
\end{eqnarray}

This action is invariant under the reparametrization transformations (\ref{reparametr}) plus $\bar{\tilde{v}}(\bar{t}) d\bar{t} = \tilde{v}(t) dt$. So, (\ref{Geo Action Lagr 1}) is the action that we need to generalize (\ref{Geo Action}). Two Lagrange multiplier are unnecessary, so we will eliminate one of them. The equation of motion for $\tilde{v}$ is:

\begin{eqnarray}
\label{v tilde} \tilde{v} = \frac{m^2 + v^{-2}\left(g_{\mu \nu} +
\kappa_2 \tilde{g}_{\mu
\nu}\right)\dot{x}^{\mu}\dot{x}^{\nu}}{2\kappa_2v^{-3}g_{\alpha
\beta}\dot{x}^{\alpha}\dot{x}^{\beta}}
\end{eqnarray}

If we now replace (\ref{v tilde}) in (\ref{Geo Action Lagr 1}), we obtain the $\tilde{\delta}$ Test Particle Action:

\begin{eqnarray}
\label{Geo Action Lagr 2} S[\dot{x},g,\tilde{g},v] = \frac{1}{4}
\int dt \left(3m^2v - v^{-1}\left(g_{\mu \nu} + \kappa_2 \tilde{g}_{\mu \nu}\right)\dot{x}^{\mu}\dot{x}^{\nu} + \frac{m^2 v^3}{g_{\alpha \beta}\dot{x}^{\alpha}\dot{x}^{\beta}}\left(m^2 + v^{-2}\left(g_{\mu \nu} + \kappa_2 \tilde{g}_{\mu \nu}\right)\dot{x}^{\mu}\dot{x}^{\nu}\right)\right)
\end{eqnarray}

The equation of motion for $v$ is still given by (\ref{v eq}). If we substitute it in (\ref{Geo Action Lagr 2}), we obtain (\ref{Geo Action}). So, (\ref{Geo Action Lagr 2}) is a good modified action to represent the trajectory of a particle in the presence of a gravitational field, given by $g$ and $\tilde{g}$, for the massive and massless case. Evaluating $m = 0$ in (\ref{Geo Action 0 Lagr}) and (\ref{Geo Action Lagr 2}), they respectively are:

\begin{eqnarray}
\label{Geo Action 0 foton} S^{(m=0)}_0[\dot{x},g,v] = -
\frac{1}{2} \int dt v^{-1}g_{\mu \nu}\dot{x}^{\mu}\dot{x}^{\nu} \\
\label{Geo Action foton} S^{(m=0)}[\dot{x},g,\tilde{g},v] = -
\frac{1}{4} \int dt v^{-1}\mathbf{g}_{\mu
\nu}\dot{x}^{\mu}\dot{x}^{\nu}
\end{eqnarray}

with $\mathbf{g}_{\mu \nu} = g_{\mu \nu} + \kappa_2 \tilde{g}_{\mu \nu}$. The equation of motion for $v$ implies that, in the usual and modified case, a massless particle will move in a null-geodesic. In the usual case we have $g_{\mu \nu}\dot{x}^{\mu}\dot{x}^{\nu} = 0$, but in our model the null-geodesic is $\mathbf{g}_{\mu \nu}\dot{x}^{\mu}\dot{x}^{\nu} = 0$.\\

On the other hand, to describe the massive case in the usual and modified models, we can use (\ref{Geo Action 0}) and (\ref{Geo Action}) respectively. We know that (\ref{Geo Action 0}) implies that a massive particle will move in a typical geodesic with $g_{\mu \nu}\dot{x}^{\mu}\dot{x}^{\nu} = -1$, after choosing $t$ equal to the proper time. On the other side, the equation of motion of the modified action is more complicated, however we can use the same gauge, $g_{\mu \nu}\dot{x}^{\mu}\dot{x}^{\nu} = -1$.\\

To conclude, in the usual case we have:

\begin{eqnarray}
\label{geodesics 0} \ddot{x}^{\mu} + \Gamma^{\mu}_{~ \alpha \beta} \dot{x}^{\alpha} \dot{x}^{\beta} = 0
\end{eqnarray}

with:

\begin{eqnarray}
\Gamma^{\mu}_{~ \alpha \beta} &=& \frac{1}{2}g^{\mu
\gamma}(g_{\gamma \alpha , \beta} + g_{\gamma
\beta , \alpha} - g_{\alpha \beta ,
\gamma}) \nonumber
\end{eqnarray}

where:

\begin{eqnarray}
\label{geodesics cond 0 m = 0}
\begin{array}{cc}
  g_{\mu \nu}\dot{x}^{\mu}\dot{x}^{\nu} = 0 & \textrm{if: } m=0
\end{array} \\
\label{geodesics cond 0 m}
\begin{array}{cc}
  g_{\mu \nu}\dot{x}^{\mu}\dot{x}^{\nu} = -1 & \textrm{if: } m>0
\end{array}
\end{eqnarray}

On the other hand, the equations of motion for a test particle in the modified case are:

\begin{eqnarray}
\label{geodesics m=0}
\begin{array}{cc} \ddot{x}^{\mu} + \mathbf{\Gamma}^{\mu}_{~ \alpha \beta} \dot{x}^{\alpha} \dot{x}^{\beta} = 0 & \textrm{if: } m=0
\end{array}\\
\label{geodesics m}
\begin{array}{cc}
M^{\mu}_{\nu}\ddot{x}^{\nu} + G^{\mu}_{\alpha \beta} \dot{x}^{\alpha} \dot{x}^{\beta} + N^{\mu}_{\alpha \beta \gamma \epsilon} \dot{x}^{\alpha} \dot{x}^{\beta} \dot{x}^{\gamma} \dot{x}^{\epsilon}  = 0 & \textrm{if: } m>0
\end{array}
\end{eqnarray}

with:

\begin{eqnarray}
\mathbf{\Gamma}^{\mu}_{~ \alpha \beta} &=& \frac{1}{2} \left[\mathbf{g}^{-1}\right]^{\mu
\gamma}(\mathbf{g}_{\gamma \alpha , \beta} + \mathbf{g}_{\gamma
\beta , \alpha} - \mathbf{g}_{\alpha \beta ,
\gamma})\nonumber \\
M^{\mu}_{\nu} &=& 2\bar{g}^{\mu}_{\nu} + \delta^{\mu}_{\nu}
\bar{g}_{\alpha \beta} \dot{x}^{\alpha} \dot{x}^{\beta} + 2
\bar{g}_{\nu \beta} \dot{x}^{\mu} \dot{x}^{\beta}
\nonumber  \\
G^{\mu}_{\alpha \beta} &=& g^{\mu \gamma}(\bar{g}_{\gamma \alpha ,
\beta} + \bar{g}_{\gamma \beta , \alpha} - \bar{g}_{\alpha \beta ,
\gamma})
\nonumber \\
N^{\mu}_{\alpha \beta \gamma \epsilon} &=& \textrm{Sym}_{\alpha
\beta \gamma \epsilon} (\bar{g}_{\beta \gamma}\Gamma^{\mu}_{~
\alpha \epsilon} + \delta^{\mu}_{\alpha}\bar{g}_{\beta \gamma ,
\epsilon}) \nonumber
\end{eqnarray}

where:

\begin{eqnarray}
\label{geodesics cond m = 0}
\begin{array}{cc}
  \mathbf{g}_{\mu \nu}\dot{x}^{\mu}\dot{x}^{\nu} = 0 & \textrm{if: } m=0
\end{array} \\
\label{geodesics cond m}
\begin{array}{cc}
  g_{\mu \nu}\dot{x}^{\mu}\dot{x}^{\nu} = -1 & \textrm{if: } m>0
\end{array}
\end{eqnarray}

This means that, in our model, massive particles do not move on geodesics of a four-dimensional metric. Only massless particles move on a null geodesic of $\mathbf{g}_{\mu \nu}$. So, $\tilde{\delta}$ gravity is not a metric theory.\\

\section{\label{Chap:Distances and time intervals.}Distances and time intervals.}

In  this section, we define the measurement of time and distances in the model.\\

The geodesic equation, (\ref{geodesics 0}), preserves the proper time of the particle along the trajectory. Eq. (\ref{geodesics m}) satisfies the same property: Along the trajectory $g_{\mu \nu}\dot{x}^{\mu}\dot{x}^{\nu}$ is constant. So, we define proper time using the original metric $g_{\mu \nu}$:

\begin{eqnarray}
\label{proper time}
d \tau = \sqrt{-g_{\mu \nu}dx^{\mu}dx^{\nu}} = \sqrt{-g_{0 0}}dx_0
\end{eqnarray}

Following \cite{Landau}, we consider the motion of light rays along infinitesimally near trajectories, using (\ref{geodesics cond m = 0}) and (\ref{proper time}), to get the three-dimensional metric:

\begin{eqnarray}
\label{tri metric}
d l^2 &=& \gamma_{i j}dx^{i}dx^{j} \\
\gamma_{i j} &=&  \frac{g_{0 0}}{\mathbf{g}_{0 0}}\left(\mathbf{g}_{i j} - \frac{\mathbf{g}_{i 0}\mathbf{g}_{j 0}}{\mathbf{g}_{0 0}}\right)\nonumber
\end{eqnarray}

Therefore, we measure proper time using the metric $g_{\mu \nu}$, but the space geometry is determined by both tensor fields, $g_{\mu \nu}$ and $\tilde{g}_{\mu \nu}$. These considerations are fundamental to explain the expansion of the universe with $\tilde{\delta}$ gravity. In the next section, we will see this in detail.\\

\section{\label{Chap:FRW and Photon Trajectory.}FRW and Photon Trajectory.}

To describe the supernova data, we must use the FRW metric. When a photon emitted from the supernova travels to  Earth, the universe is expanding. This means that the photon is affected by the cosmological Doppler effect. So the metric $g$ is:

\begin{eqnarray}
\label{g FRW} g_{\mu \nu} =
\left(%
\begin{array}{cccc}
  - c^2 &    0   &     0     &          0              \\
   0 & R^2(t) &     0     &          0              \\
   0 &    0   & R^2(t)r^2 &          0              \\
   0 &    0   &     0     & R^2(t)r^2\sin^2(\theta) \\
\end{array}%
\right)
\end{eqnarray}

where $R(t)$ is the scale factor that depends on the time parameter $t$. Assuming an isotropic and homogeneous universe, we can use the following ansatz for $\tilde{g}$:

\begin{eqnarray}
\label{tg FRW} \tilde{g}_{\mu \nu} =
\left(%
\begin{array}{cccc}
  - c^2 \tilde{A}(t) &       0       &        0         &             0                  \\
        0        & \tilde{B}(t) &        0         &             0                  \\
        0        &       0       & \tilde{B}(t)r^2 &             0                  \\
        0        &       0       &        0         & \tilde{B}(t)r^2\sin^2(\theta) \\
\end{array}%
\right)
\end{eqnarray}

So, $\tilde{g}$ only has two independent functions, $\tilde{A}(t)$ and $\tilde{B}(t)$, that depend of the time parameter $t$ just like $R(t)$. We know that the Einstein's equations lead to:

\begin{eqnarray}
\label{Eq Eins} \left(\frac{\dot{R}(t)}{R(t)}\right)^2 &=&
\frac{\kappa}{3} \sum_i \rho_i(t) \\
\label{Eq Cont} \dot{\rho}_i(t) &=& -
\frac{3\dot{R}(t)}{R(t)}(\rho_i(t) + p_i(t))
\end{eqnarray}

with $\dot{f}(t) = \frac{d f}{d t}(t)$. But now we have additional equations arising from those of $\tilde{g}_{\mu \nu}$ given by (\ref{tilde Eq 2}). These equations are:

\begin{eqnarray}
\label{Eq Eins tild 1} \frac{\dot{R}(t)}{R(t)}\dot{\tilde{B}}(t) -
\frac{\dot{R}^2(t)}{R^2(t)}\tilde{B}(t) -
\frac{\ddot{R}(t)}{R(t)}\tilde{B}(t) - \dot{R}^2(t)\tilde{A}(t) &=& 0\\
\label{Eq Eins tild 2} \ddot{\tilde{B}}(t) - \frac{\dot{R}(t)}{R(t)}\dot{\tilde{B}}(t) -
2\frac{\ddot{R}(t)}{R(t)}\tilde{B}(t) - R(t)\dot{R}(t)\dot{\tilde{A}}(t) -
R(t)\ddot{R}(t)\tilde{A}(t) - 2\dot{R}^2(t)\tilde{A}(t) &=& 0
\end{eqnarray}

To solve the system (\ref{Eq Eins}-\ref{Eq Eins tild 2}), we need equations of state which relate $\rho_i(t)$ and $p_i(t)$, for which we take $p_i(t) = \omega_i\rho_i(t)$. Since we wish to explain DE with $\tilde{\delta}$ gravity, we will assume that the universe  only have non relativistic matter (cold dark matter, baryonic matter) and radiation (photons, massless particles). So, we will require two equations of state. For non relativistic matter we use $p_M(t) = 0$ and for radiation $p_R(t) = \frac{1}{3}\rho_R(t)$, where we have assumed that their interaction is null. Replacing in (\ref{Eq Eins})-(\ref{Eq Eins tild 2}) and solving them, we find the exact solution:

\begin{eqnarray}
\rho(t) &=& \rho_M(t) + \rho_R(t) \nonumber \\
\label{rho(t)}&=& \frac{H_0^2 \Omega_R}{\kappa C^4} \frac{X(t) + 1}{ X^4(t)} \\
p(t) &=& \frac{1}{3} \rho_R(t) \nonumber \\
\label{p(t)}&=& \frac{H_0^2 \Omega_R}{3 \kappa C^4} \frac{1}{X^4(t)} \\
\label{Sol X(t)}
\sqrt{X(t) + 1}\left(X(t) - 2\right) &=& \frac{3 H_0\sqrt{\Omega_R}}{2 C^2} t - 2 \\
\label{Sol tA(t)}
\tilde{A}(t) &=& - \frac{\bar{l}}{\kappa_2 C^{\frac{5}{2}}}\frac{\sqrt{X(t)+1}}{X(t)\left(3 X(t)+4\right)^2} \\
\label{Sol tB(t)}
\tilde{B}(t) &=&
- R_{eq}^2\frac{\bar{l}}{4 \kappa_2 C^{\frac{5}{2}}}\frac{X(t)\sqrt{X(t)+1}}{3 X(t)+4} \\
X(t) &=& \frac{R(t)}{R_{eq}}
\end{eqnarray}

Where $R_{eq}$ is the scale factor at matter-radiation equality, that is $\rho_M(t_{eq}) = \rho_R(t_{eq})$, $\bar{l}$ is an arbitrary parameter, $C = \frac{\Omega_R}{\Omega_M}$, and $\Omega_R$ and $\Omega_M$ are the radiation and non relativistic matter density in the present respectively, with $\Omega_M = 1 - \Omega_R$. We know that $\Omega_R \ll 1$, so $\Omega_M \sim 1$ and $C \ll 1$. By definition $X(t) \gg 1$ describes the non relativistic era and $X(t) \ll 1$ describes the radiation era.\\

Since we have the cosmological solution of the $\tilde{\delta}$ gravity Action now, we can analyze the trajectory of a supernova photon when it is traveling to  Earth. For this we use (\ref{geodesics cond m = 0}) in a radial trajectory from $r_1$ to $r=0$. So, we have:

\begin{eqnarray}
- (1+\kappa_2 \tilde{A}(t)) c^2 dt^2 + (R^2(t)+\kappa_2 \tilde{B}(t)) dr^2 = 0 \nonumber
\end{eqnarray}

In the usual case, we have that $c dt = - R(t) dr$. In the $\tilde{\delta}$ gravity case, we define the modified scale factor:

\begin{eqnarray}
\label{R tilde}
\tilde{R}(t) = \sqrt{\frac{R^2(t)+\kappa_2 \tilde{B}(t)}{1+\kappa_2 \tilde{A}(t)}}
\end{eqnarray}

such that $c dt = - \tilde{R}(t) dr$ now. With this definition, we obtain that:

\begin{eqnarray}
\label{r1 1}
r_1 = c \int_{t_1}^{t_0} \frac{dt}{\tilde{R}(t)}
\end{eqnarray}

If a second wave crest is emitted at $t = t_1 + \Delta t_1$ from $r = r_1$, it will reach $r = 0$ at $t = t_0 + \Delta t_0$, so:

\begin{eqnarray}
\label{r1 2}
r_1 = c \int_{t_1 + \Delta t_1}^{t_0 + \Delta t_0} \frac{dt}{\tilde{R}(t)}
\end{eqnarray}

Therefore, for $\Delta t_1$, $\Delta t_0$ small, which is appropriate for light waves, we get:

\begin{eqnarray}
\frac{\Delta t_0}{\Delta t_1} = \frac{\tilde{R}(t_0)}{\tilde{R}(t_1)}
\end{eqnarray}

Since $t$ is the proper time according to (\ref{proper time}), we have that

\begin{eqnarray}
\frac{\Delta \nu_1}{\Delta \nu_0} = \frac{\tilde{R}(t_0)}{\tilde{R}(t_1)}
\end{eqnarray}

where $\nu_0$ is the light frequency detected at $r = 0$, corresponding to a source emission at frequency $\nu_1$. So, the redshift is now:

\begin{eqnarray}
\label{redshift}
1 + z(t_1) = \frac{\tilde{R}\left(t_0\right)}{\tilde{R}(t_1)}
\end{eqnarray}

We see that $\tilde{R}\left(t\right)$ replaces the usual scale factor $R(t)$ in the calculation of $z$. This means that we need to redefine the luminosity distance too. For this, let us consider a mirror of radius $b$ that is receiving light from our distant source at $r_1$. The photons that reach the mirror are inside a cone of half-angle $\epsilon$ with origin at the source.\\

Let us compute $\epsilon$. The path of the light rays is given by $\vec{r}(\rho) = \rho \hat{n} + \vec{r}_1$, where $\rho > 0$ is a parameter and $\hat{n}$ is the direction of the light ray. Since the mirror is in $\vec{r} = 0$, then $\rho = r_1$ and $\hat{n} = - \hat{r}_1 + \vec{\epsilon}$, where $\epsilon$ is the angle between $-\vec{r}_1$ and $\hat{n}$ at the source, forming a cone. The proper distance is determined by the 3-dimensional metric (\ref{tri metric}), so we get $b =  \tilde{R}(t_0) r_1 \epsilon$. Then, the solid angle of the cone is:

\begin{eqnarray}
\Delta \Omega &=& \int_0^{2 \pi} d\phi \int_0^{\epsilon} \sin(\theta) d\theta = 2\pi (1-\cos(\epsilon)) \nonumber \\
&=& \pi \epsilon^2 = \frac{A}{r_1^2 \tilde{R}^2(t_0)} \nonumber
\end{eqnarray}

where $A = \pi b^2$ is the proper area of the mirror. This means that $\epsilon = \frac{b}{r_1 \tilde{R}(t_0)}$. So, the fraction of all isotropically emitted photons that reach the mirror is:

\begin{eqnarray}
f &=& \frac{\Delta \Omega}{4 \pi} \nonumber \\
&=& \frac{A}{4 \pi r_1^2 \tilde{R}^2(t_0)} \nonumber
\end{eqnarray}

We know that the apparent luminosity, $l$, is the received power per unit mirror area. Power is energy per unit time, so the received power is $P = \frac{h\nu_0}{\Delta t_0} f$, where $h\nu_0$ is the energy corresponding to the received photon, and the total emitted power by the source is $L = \frac{h\nu_1}{\Delta t_1}$, where $h\nu_1$ is the energy corresponding to the emitted photon. Therefore, we have that:

\begin{eqnarray}
P &=& \frac{\tilde{R}^2(t_1)}{\tilde{R}^2(t_0)} L f \nonumber \\
l &=& \frac{P}{A} \nonumber \\
&=& L \frac{\tilde{R}^2(t_1)}{\tilde{R}^2(t_0)} \frac{1}{4 \pi r_1^2 \tilde{R}^2(t_0)}  \nonumber
\end{eqnarray}

where we have used that $\frac{\Delta t_0}{\Delta t_1} = \frac{\nu_1}{\nu_0} = \frac{\tilde{R}(t_0)}{\tilde{R}(t_1)}$. On the other hand, we know that, in a Euclidean space, the luminosity decreases with distance $d_L$ according to $l = \frac{L}{4\pi d_L^2}$. Therefore, using (\ref{r1 1}), the luminosity distance is:

\begin{eqnarray}
d_L &=& \frac{\tilde{R}^2(t_0)}{\tilde{R}(t_1)} r_1 \nonumber \\
&=& c \frac{\tilde{R}^2\left(t_0\right)}{\tilde{R}(t_1)} \int_{t_1}^{t_0} \frac{dt}{\tilde{R}(t)}
\end{eqnarray}

Moreover, we can use (\ref{Sol X(t)}) to change the $t$ variable for $Y = C X(t) = \frac{R(t)}{R\left(t_0\right)}$ (the scale factor normalized to one in the present), and then define $\tilde{Y} = \frac{\tilde{R}(t)}{R\left(t_0\right)}$. Using (\ref{Sol tA(t)}) and (\ref{Sol tB(t)}) in (\ref{R tilde}), we obtain:

\begin{eqnarray}
\label{Y tilde}
\tilde{Y}[Y,C,\bar{l}] = Y \sqrt{\frac{(3Y+4C) \left(4 C Y(3Y+4C) - \bar{l} \sqrt{Y+C}\right)}{4 C \left(Y(3Y+4C)^2 - \bar{l}\sqrt{Y+C}\right)}}
\end{eqnarray}

We notice that $\tilde{Y} \sim Y$ in the radiation era, that is $Y \ll C$, so the universe evolves normally in the beginning of the universe, without differences with GR. But when $Y \gg C$, $\tilde{Y} \gg Y$ and we will have a Big Rip, i.e. $\tilde{Y} = \infty$, when the denominator is null. That is $Y \sim \left(\frac{\bar{l}}{9}\right)^{\frac{2}{5}}$. We will give more detail for this when we will study the supernova data. Now, with all our definitions, the luminosity distance is reduced to:

\begin{eqnarray}
\label{d_L 0}
d_L =  c \frac{\sqrt{C}}{H_0\sqrt{\Omega_R}} \frac{\tilde{Y}_0^2[C,\bar{l}]}{\tilde{Y}[Y,C,\bar{l}]} \int_{Y}^{1} \frac{Y' dY'}{\tilde{Y}[Y',C,\bar{l}] \sqrt{Y'+C}}
\end{eqnarray}

with $\tilde{Y}_0[C,\bar{l}] = \tilde{Y}[1,C,\bar{l}]$. This means that the distances will be different now. 

In GR, we have:

\begin{eqnarray}
\label{d_L 0 usual}
d_L = \frac{c}{Y H_0} \int_{Y}^{1} \frac{dY'}{\sqrt{\Omega_{\Lambda} Y'^4 + \Omega_M Y' + \Omega_R}}
\end{eqnarray}

where $\Omega_{\Lambda} = 1 - \Omega_M - \Omega_R$ is the dark energy density in the present. We will use (\ref{d_L 0 usual}) to compare both, a universe with dark energy and our modified gravitation model, with the supernova data.\\

Finally, we note that (\ref{Sol X(t)}) gives us the time coordinate. In the new notation, it is:

\begin{eqnarray}
\label{t}
t = \frac{2 C^{\frac{1}{2}}}{3 H_0\sqrt{\Omega_R}}\left(\sqrt{Y + C}\left(Y - 2 C\right) + 2 C^{\frac{3}{2}}\right)
\end{eqnarray}

Therefore, it is possible to obtain a different age of the universe. A different perception of the distances implies a different perception of time.All these differences arise a consequence of the modified trajectory of photons.\\

\section{\label{Chap:Analysis and Results.}Analysis and Results.}

Before we analyze the data, we will define the parameters to be determined. In GR, $d_L$ depends upon four parameters: $Y$, $H_0 = 100 h$ km s$^{-1}$ Mpc$^{-1}$, $\Omega_M$ and $\Omega_R$ according to (\ref{d_L 0 usual}). However, the CMB black body spectrum give us the photons density in the present, $\Omega_{\gamma}$, and if we assume that $\Omega_R = \Omega_{\gamma} + \Omega_{\nu} = \left(1+3 \left(\frac{7}{8}\right)\left(\frac{4}{11}\right)^{4/3}\right) \Omega_{\gamma}$, we obtain $h^2\Omega_R = 4.15 \times 10^{-5}$. Therefore, the parameters in $d_L$ can be reduced to three: $Y$, $h$ and $h^2\Omega_M$. For the same reasons, in our modified gravity, $d_L$ depends only on three parameters: $Y$, $C$ and $\bar{l}$, as shown in (\ref{d_L 0}). We use $H_0 \sqrt{\Omega_R} = 0.644$ km s$^{-1}$ Mpc$^{-1}$\\

The supernova data gives the apparent magnitude as a function of redshift. For this reason, it is useful to use $z$ instead of $Y$. So, we have:\\

In GR:

\begin{eqnarray}
\label{m usual}
m[z,h,h^2\Omega_M] &=& M + 5\log_{10}\left(\frac{d_L[z,h,h^2\Omega_M]}{10 \textrm{ pc}}\right) \\
\label{d_L usual}
d_L[z,h,h^2\Omega_M] &=&  \frac{c (1+z) \textrm{ Mpc s}}{100 \textrm{ km}} \int_{\frac{1}{1 + z}}^{1} \frac{dY'}{\sqrt{(h^2 - h^2\Omega_M - h^2\Omega_R) Y'^4 + h^2\Omega_M Y' + h^2\Omega_R}} \\
\label{z usual}
Y(z) &=& \frac{1}{1 + z}
\end{eqnarray}

In our modified gravity:

\begin{eqnarray}
\label{m}
m[z,C,\bar{l}] &=& M + 5 \log_{10}\left(\frac{d_L[z,C,\bar{l}]}{10 \textrm{ pc}}\right) \\
\label{d_L}
d_L[z,C,\bar{l}] &=&  c (1+z) \frac{\sqrt{C}\tilde{Y}_0[C,\bar{l}]}{H_0\sqrt{\Omega_R}} \int_{1}^{Y(z)} \frac{Y' dY'}{\tilde{Y}[Y',C,\bar{l}] \sqrt{Y'+C}} \\
\label{z}
\tilde{Y}[Y(z),C,\bar{l}] &=& \frac{\tilde{Y}_0[C,\bar{l}]}{1 + z}
\end{eqnarray}

where $m$ is the apparent magnitude and $M$ is the absolute magnitude, common to all supernova, so it is constant. To find $Y(z)$, we must solve (\ref{z}) using (\ref{Y tilde}), numerically. Now we will introduce the statistical method to fit the data.\\

We interpret errors in data by the variance $\sigma$ in a normally distributed random variable. If we are fitting a function $y(x)$ to a set of points $(x_i, y_i)$ with errors $(\sigma_{x i}, \sigma_{y i})$, we must minimize \cite{Numeric}:

\begin{eqnarray}
\chi^2(\textrm{per point}) &=& \frac{1}{N}\sum^N_{i=1} \frac{(y_i - y (x_i))^2}{\sigma_{f i}^2} \nonumber \\
\sigma_{f i}^2 &=& \sigma_{y i}^2 + y' (x_i)^2 \sigma_{x i}^2 \nonumber
\end{eqnarray}

In our case, we want to fit the data $(z_i, m_i)$ with errors $(\sigma_{z i}, \sigma_{m i})$ to the model:

\begin{eqnarray}
m(z) = M + 5 \log_{10} \left(\frac{d_L(z)}{10 pc}\right) \nonumber
\end{eqnarray}

Therefore, we must minimize:

\begin{eqnarray}
\label{chi square}
\chi^2(\textrm{per point}) = \frac{1}{N}\sum^N_{i=1} \frac{(m_i - m(z_i))^2}{\sigma_{m i}^2 + \left(\frac{d m}{d z} (z_i)\right)^2 \sigma_{z i}^2}
\end{eqnarray}

Now, we can proceed to analyze the supernova data given in \cite{Data}. In both cases, $d_L$ is given by an exact expression, but we need to use a numeric method to solve the integral and fit the data to determinate the optimum values for the parameters that represent the m v/s z of the supernova data. Minimizing (\ref{chi square}), we obtain:\\

In GR $h = 0.6603 \pm 0.0078$ and $h^2\Omega_M = 0.096 \pm 0.014$ with $\chi^2(\textrm{per point}) = 1.033$.\\

In our modified gravity $\bar{l} = 34.42 \pm 4.27$ and $C = (3.61 \pm 0.21)\times 10^{-4}$ with $\chi^2(\textrm{per point}) = 1.029$.\\

With these values, we can calculate the age of the universe. We know that, in the usual case, it is $1.37 \times 10^{10}$ years, but that in our model it is given by (\ref{t}). Substituting  the corresponding values for $\bar{l}$, $C$ and taking $Y = 1$, we obtain $1.92 \times 10^{10}$ years. Finally, we can calculate when the Big-Rip will happen. For this, we need $Y$ when $\tilde{Y} = \infty$. That is $Y = 1.71$, so $t_{\textrm{Big Rip}} = 4.3 \times 10^{10}$ years. Therefore, the universe almost has lived half of its life. Phantom fields also produce a cosmological model that have this property \cite{Phantom 1,Phantom 2}. In the\ $\tilde{\delta}$ gravity model we can avoid a Big Rip at later time by a mechanism that give masses to all massless particles. Some options are quantum effects (which are finite in this model) or massive photons due to superconductivity \cite{SuperConduct} which could happen at very low temperatures, which are natural at a later stages of the expansion of the Universe.\\

\section{\label{Chap: Dark Matter.}Dark Matter.}

The principal objective of this paper is to explain the accelerated expansion of the universe without dark energy, using a modified gravity model. But it is possible that this model explain dark matter too. In this section we present a preliminary discussion of this important topic.\\

Far from a source, the gravitational field corresponds to a point-like source and we have a Schwarzschild solution for $g_{\mu \nu}$. With this, we can get a solution to $\tilde{g}_{\mu \nu}$ using (\ref{tilde Eq 2}). So, we have:

\begin{eqnarray}
g_{\mu \nu} &=& \left( \begin{array}{cccc}
        - c^2 \left(1 - \frac{2 GM}{c^2 r}\right) & 0 & 0 & 0\\
        0 & \left(1 - \frac{2 GM}{c^2 r}\right)^{-1} & 0 & 0\\
        0 & 0 & r^2 & 0\\
        0 & 0 & 0 & r^2 \sin^2(\theta)
      \end{array} \right) \nonumber \\
\tilde{g}_{\mu \nu} &=& \left( \begin{array}{cccc}
        - c^2 \left (1 - \frac{2 GM (1-b)}{c^2 r}\right) & 0 & 0 & 0\\
        0 & \frac{1 - \frac{2 GM (1+b)}{c^2 r}}{\left(1 - \frac{2 GM}{c^2 r}\right)^2} & 0
        & 0\\
        0 & 0 & r^2 & 0\\
        0 & 0 & 0 & r^2 \sin^2(\theta)
      \end{array} \right) \nonumber
\end{eqnarray}

where $b$ is a new parameter. We have imposed the boundary condition $g_{\mu \nu} \sim \eta_{\mu \nu}$ and $\tilde{g}^{\mu \nu} \sim \eta^{\mu \nu}$ for $r \rightarrow \infty$. On the other hand, if we define $h_{\mu \nu} = g_{\mu \nu} - \eta_{\mu \nu}$ and $\tilde{h}_{\mu \nu} = \tilde{g}_{\mu \nu} - \eta_{\mu \nu}$, and we use the equation of motion for a test particle (\ref{geodesics m}), we obtain that:

\begin{eqnarray}
\ddot{\vec{x}} &=& -\nabla \phi \nonumber \\
\phi &=& - \frac{1}{2}\left(\frac{(2 - \kappa_2) h_{00} + 2 \kappa_2 \tilde{h}_{00}}{2 + \kappa_2}\right) \nonumber
\end{eqnarray}

Where $\vec{x}$ is the space vector and $\phi$ is the potential. In GR $\phi = -\frac{h_{00}}{2}$, therefore $\phi = - \frac{GM}{r}$ for $r \rightarrow \infty$. So, in our model, we will have, for $r\rightarrow \infty$, that:

\begin{eqnarray}
\phi &=& - \frac{GM}{r} \left(1 - \frac{\kappa_2 b}{1 + \frac{\kappa_2}{2}}\right) \nonumber
\end{eqnarray}

Then, if we define the detected total mass, $M_T$, such that $\phi = - \frac{GM_T}{r}$, we will get:

\begin{eqnarray}
M_T &=& M + M_{DM} \\
M_{DM} &=& - \frac{\kappa_2 b M}{1 + \frac{\kappa_2}{2}}
\end{eqnarray}

where $M_{DM}$ can be interpreted as the dark matter. This means that, with $b<0$, we could explain dark matter. We can calculate the dark matter with photons too. For this, we use a radial trajectory in a null geodesic, given by (\ref{geodesics cond m = 0}), for $r \rightarrow \infty$. With this, we obtain:

\begin{eqnarray}
- c^2 \left(1 + \kappa_2 - \frac{2 M (1+\kappa_2 (1-b))}{r}\right) dt^2 + \left(1 + \kappa_2 + \frac{2 M (1+\kappa_2 (1-b))}{r}\right) dr^2 = 0
\end{eqnarray}

Comparing with the usual case where:

\begin{eqnarray}
- c^2 \left(1 - \frac{2 M}{r}\right) dt^2 + \left(1 + \frac{2 M}{r}\right) dr^2 = 0
\end{eqnarray}

We obtain that $M_T = \frac{M (1+\kappa_2 (1-b))}{1 + \kappa_2}$. Therefore:

\begin{eqnarray}
M_{DM} = - \frac{\kappa_2 b M}{1 + \kappa_2}
\end{eqnarray}

We notice that photons and massive particles see different $M_{DM}$, but since $\kappa_2$ is very small, this difference is hard to detect. This means that (\ref{Geo Action}) and (\ref{Geo Action Lagr 2}) is a consistent system to describe dark matter. These preliminary results provide an excellent motivation to explain dark matter with this modified model.\\

\section*{CONCLUSIONS.}

We have proposed a modified gravity model with good properties at the quantum level. It is finite on shell in the vacuum and only lives at one loop. It incorporates a new field $\tilde{g}_{\mu \nu}$ that transforms correctly under general coordinate transformation and exhibits a new symmetry: the $\tilde{\delta}$ symmetry. The new action is invariant under these transformations. We call this new gravity model $\tilde{\delta}$ gravity. A quantum field theory analysis of $\tilde{\delta}$ gravity has been developed \cite{delta gravity,DG DE}.\\

In this paper, we study the classical effects in a cosmological setting. To this end, we require to set up the following two issues. First, we need to find the equations for $\tilde{\delta}$ gravity. One of them is Einstein's equation, which it gives us $g_{\mu \nu}$, and the other equation is (\ref{tilde Eq}) to solve for $\tilde{g}_{\mu \nu}$. Second, we need the modified test particle action. This action, (\ref{Geo Action Lagr 2}), incorporates the new field $\tilde{g}_{\mu \nu}$. We obtain that a photon, or a massless particle, moves in a null geodesic of $\mathbf{g}_{\mu \nu} = g_{\mu \nu} + \kappa_2 \tilde{g}_{\mu \nu}$ and that a massive particle is governed by the equation of motion (\ref{geodesics m}). With all this basic set up, we can study any cosmological phenomenon.\\

In \cite{DG DE} it was shown that $\tilde{\delta}$ gravity predicts an accelerated expansion of the universe without a cosmological constant or additional scalar fields by using an approximation corresponding to small redshift. We find in the present work an exact expression for the cosmological luminosity distance and verify that $\tilde{\delta}$ gravity do not require dark energy. With this exact expression, we could also study very early phenomenon in the universe, for example inflation and the CMB power spectrum. This work is in progress.\\

On the other hand, photons move on a null geodesic of $\mathbf{g}$, so we can define a new scale factor $\tilde{R}(t)$. If we assume that the universe only has non relativistic matter and radiation, we can obtain an exact expression for $\tilde{R}(t)$. It is clear in (\ref{Y tilde}) that $1 \gg C \neq 0$ is necessary to obtain an accelerated expansion of the universe. Therefore a minimal component of radiation explains the supernova data without dark energy. In this way, in this model, the accelerated expansion of the universe, can be understood as a geometric effect.\\

We also calculate the age of the universe. We find that the universe has lived a bit more as in GR. This is not a contradiction, but rather a reinterpretation of the observations. This result is a consequence of the new equation of motion for the photons. This model ends in a Big Rip and we calculate when it will happen. The universe almost has lived half of its life. Even though the Big Rip could be seen as as a problem, we observe that other cosmological models share this property too \cite{Phantom 1,Phantom 2}. Nevertheless, in our case, we have some way outs from the Big Rip. For example, the appearance of quantum effects or massive photons at times close to the Big Rip, by effects similar to superconductivity \cite{SuperConduct}. These effects could happen at very low temperatures which are common at the later stages of the evolution of the Universe.\\

Finally, we introduce a preliminary scenario to explain dark matter with $\tilde{\delta}$ gravity. We find that, if we assume a point-like source, the Schwarzschild solution is obtained for $g$ and a similar solution for $\tilde{g}$ but with extra parameter, $b$. If we study the asymptotic region, we find that an extra mass is perceived by a test particle, massive or massless. This extra mass is determined by $b$ and could be understood as a virtual dark matter. To determine whether or not this is a solution of the dark matter problem we have to study the motion of stars around the center of the host galaxy. Such a study is been pursued now but goes beyond the scope of the present paper.\\

\section*{Acknowledgements.}

PG acknowledges support from Beca Doctoral Conicyt $\#$ 21080490.
The work of JA is partially supported by Fondecyt 1110378. The
authors especially thank L.F. Urrutia for carefully reading of the manuscript.


\begin{thebibliography}{99}

\bibitem{GR scale} Clifford M. Will, Living Rev. Relativity 9 (2006), http://www.livingreviews.org/lrr-2006-3.

\bibitem{string 1} M.B. Green, J.H. Schwarz, E. Witten. \textit{Superstring Theory.} Vols. 1, 2, Cambridge University Press, 1987.

\bibitem{string 2} J. Polchinski. \textit{String Theory.} Vols. 1, 2, Cambridge University Press, 1998.

\bibitem{Weinberg} S. Weinberg. \textit{Cosmology.} Oxford University Press, 2008.

\bibitem{DM DE 1} D. Hooper, E.A. Baltz. Annu. Rev. Nucl. Part. Sci. 58 (2008) 293.

\bibitem{DM DE 2} A.G. Riess, et al. Astron. J. 116 (1998) 1009.

\bibitem{DM DE 3} S. Perlmutter, et al. Astrophys. J. 517 (1999) 565.

\bibitem{DM DE 4} R.R. Caldwell, M. Kamionkowski. Annu. Rev. Nucl. Part. Sci. 59 (2009) 397.

\bibitem{DM DE 5} J.A. Frieman, M.S. Turner, D. Huterer. Annu. Rev. Astron. Astrophys. 46 (2008) 385.

\bibitem{DM DE 6} M. Milgrom, Astrophys. J. 270 (1983) 365.

\bibitem{DM DE 7} J. Bekenstein, Phys. Rev. D 70 (2004) 083509.

\bibitem{DE 1} A. Albrecht, et al. arXiv:astro-ph/0609591, (2006).

\bibitem{DE 2} J.A. Peacock, et al. ESA-ESO Working Groups Report No. $3$. arXiv:astro-ph/0610906, (2006).

\bibitem{DE 3} S. Tsujikawa, Lecture Notes in Physics. 800 (2010) 99.

\bibitem{GR Weinberg} S. Weinberg, in: S.W. Hawking, W. Israel (Eds.). \textit{General Relativity: An Einstein Centenary Survey.} Cambridge University Press, 1979, Chapter 16, p. 790.

\bibitem{induced gravity 1} Ya.B. Zeldovich, JETP Lett. 6 (1967) 316.

\bibitem{induced gravity 2} A. Sakharov, Sov. Phys. Dokl. 12 (1968) 1040.

\bibitem{induced gravity 3} O. Klein, Phys. Scr. 9 (1974) 69.

\bibitem{induced gravity 4} S. Adler, Rev. Mod. Phys. 54 (1982) 729.

\bibitem{Ren group 1} D.F. Litim. Phys. Rev. Lett. 92 (2004) 201.

\bibitem{Ren group 2} D.F. Litim. JHEP07(2005)005 doi:10.1088/1126-6708/2005/07/005.

\bibitem{Ren group 3} A. Codello, R. Percacci, C. Rahmede. Annals Phys. 324 (2009) 414.

\bibitem{Ren group 4} M. Reuter, F. Saueressig. Lectures given at First Quantum Geometry and Quantum Gravity School, Zakopane, Poland. arXiv:0708.1317, 2007.

\bibitem{Lorentz triang} J. Ambjorn, J. Jurkiewicz, R. Loll, Phys. Rev. Lett. 85 (2000) 924.

\bibitem{Alfaro 0} J. Alfaro, D. Espriu, D. Puigdomenech. Phys. Rev. D 86 (2012) 025015.

\bibitem{Alfaro 1} J. Alfaro, D. Espriu, D. Puigdomenech. Phys. Rev. D 82 (2010) 045018.

\bibitem{work 1} C.J. Isham, A. Salam, J.A. Strathdee, Annals Phys. 62 (1971) 98.

\bibitem{work 2} A.B. Borisov, V.I. Ogievetsky, Theor. Math. Phys. 21 (1975) 1179.

\bibitem{work 3} E.A. Ivanov, V.I. Ogievetsky, Lett. Math. Phys. 1 (1976) 309.

\bibitem{work 4} D. Amati, J. Russo, Phys. Lett. B 248 (1990) 44.

\bibitem{work 5} J. Russo, Phys. Lett. B 254 (1991) 61.

\bibitem{work 6} A. Hebecker, C. Wetterich, Phys. Lett. B 574 (2003) 269.

\bibitem{work 7} C. Wetterich, Phys. Rev. D 70 (2004) 105004.

\bibitem{tHooft} G. 't Hooft, M. Veltman. Ann. Inst. Henri Poincar\'e 20 (1974) 69.

\bibitem{Alfaro 2} J. Alfaro. arXiv:hep-th/9702060.

\bibitem{Alfaro 3} J. Alfaro, P. Labra\~na. Phys. Rev. D 65 (2002) 045002.

\bibitem{delta gravity} J. Alfaro, P. Gonzalez, R. Avila. Class. Quant. Grav. 28 (2011) 215020.

\bibitem{DG DE} J. Alfaro. Physics Letters B. 709 (2012) 101-105.

\bibitem{Phantom 1} R.R. Caldwell, M. Kamionkowski, N.N. Weinberg, Phys. Rev. Lett. 91 (2003) 071301.

\bibitem{Phantom 2} R.R. Caldwell, Phys. Lett. B 545 (2002) 2329.

\bibitem{Landau} L. Landau, L.M. Lifshitz. \textit{The Classical Theory of Fields.} Pergamon Press, 1980.

\bibitem{Numeric} W. Press, S. Teukolsky, W. Vetterling, B. Flannery. \textit{Numerical Recipes in C.} Second edition. Cambridge University Press, 1992.

\bibitem{Data} W.M. Wood-Vasey, et al., Astrophys. J. 666 (2007) 674.

\bibitem{SuperConduct} B. Sakita. \textit{Quantum Theory of Many Variable Systems and Fields.} World Scientific Lecture Notes in Physics, Vol. 1. Chap 5.

\end{thebibliography}
\end{document}